\pdfoutput=1

\documentclass[
  reprint,
  superscriptaddress,
 amsmath,amssymb,
 aps,prl
]{revtex4-1}

\usepackage{graphicx}
\usepackage{dcolumn}
\usepackage{bm}
\usepackage{color}

\bmdefine{\bdi}{i}
\bmdefine{\bdj}{j}
\bmdefine{\bdx}{x}
\bmdefine{\bdy}{y}
\bmdefine{\bdr}{r}
\bmdefine{\bdR}{R}
\bmdefine{\bdS}{S}
\bmdefine{\bdL}{L}
\bmdefine{\bdJ}{J}
\bmdefine{\bdA}{A}
\bmdefine{\bdE}{E}
\bmdefine{\bdD}{D}
\bmdefine{\bdQ}{Q}
\bmdefine{\bdq}{q}
\bmdefine{\bdzero}{0}
\bmdefine{\bddelta}{\delta}

\begin{document}

\title{
  Floquet engineering of Mott insulators
  with strong spin-orbit coupling
}

\author{Naoya Arakawa}
\email{arakawa@phys.chuo-u.ac.jp}
\affiliation{The Institute of Science and Engineering,
  Chuo University, Bunkyo, Tokyo 112-8551, Japan}
\author{Kenji Yonemitsu}
\affiliation{The Institute of Science and Engineering,
  Chuo University, Bunkyo, Tokyo 112-8551, Japan}
\affiliation{Department of Physics,
  Chuo University, Bunkyo, Tokyo 112-8551, Japan}

\date{\today}

\begin{abstract}
  We propose a method for controlling the exchange interactions
  of Mott insulators with strong spin-orbit coupling.
  We consider a multiorbital system with strong spin-orbit coupling
  and a circularly polarized light field
  and derive its effective Hamiltonian in the strong-interaction limit.
  Applying this theory to a minimal model of $\alpha$-RuCl$_{3}$,
  we show that
  the magnitudes and signs of 
  three exchange interactions, $J$, $K$, and $\Gamma$,
  can be changed simultaneously.
  Then,
  considering another case in which one of the hopping integrals has a different value
  and the other parameters are the same as those for $\alpha$-RuCl$_{3}$,
  we show that
  the Heisenberg interaction $J$ can be made much smaller than
  the anisotropic exchange interactions $K$ and $\Gamma$.

\end{abstract}
\maketitle


Periodic driving enables us to control the magnetic properties of solids.
The solution to the Schr\"{o}dinger equation
for a periodically driven system
satisfies the Floquet theorem
because of time periodicity of the driving field~\cite{Floquet1,Floquet2}.
In particular,
the time evolution in steps of the driving period $T$
can be described by a time-independent Hamiltonian~\cite{Kuwahara}.
Since its parameters usually depend on
the amplitude and frequency of the driving field,
the properties could be controlled by tuning the driving field;
such control is called Floquet engineering~\cite{review1,review2,review3}. 
For example,
by applying $E(t)=E_{0}\cos \omega t$ 
to a single-orbital Mott insulator and tuning $E_{0}$ and $\omega$, 
we can change the magnitude and sign of 
the antiferromagnetic Heisenberg interaction
between magnetic moments~\cite{Floquet-1orbMott1,Floquet-1orbMott2}.
Moreover,
for a multiorbital Mott insulator without spin-orbit coupling (SOC),
we can control 
the antiferromagnetic and the ferromagnetic contributions
to the Heisenberg interaction 
via a time-periodic electric field~\cite{Floquet-MultiMott1,Floquet-MultiMott2}.
Such control could be used to engineer the magnetic properties of solids 
because the exchange interactions 
are key quantities describing
magnetization dynamics~\cite{review-dynamics}
and spintronics phenomena~\cite{review-spintronics}.

Although the magnetic properties of solids are 
affected by the SOC,
Floquet engineering of Mott insulators with strong SOC is lacking.
The magnetic properties of Mott insulators with strong SOC
are described by
spin and orbital coupled degrees of freedom~\cite{StrongLS-Mott,StrongLS-Mott-review}.
As a result,
not only the Heisenberg interaction 
but also the anisotropic exchange interactions 
contribute to
the effective
Hamiltonian~\cite{StrongLS-Superex1,StrongLS-Superex2,StrongLS-Superex3,Rau-PRL,NA}. 
For example,
the effective Hamiltonians for $\alpha$-RuCl$_{3}$ and the honeycomb iridates 
possess the Heisenberg interaction, 
the Kitaev interaction, and
the off-diagonal symmetric
exchange interaction~\cite{Rau-PRL,Rau-review,Valenti-PRB,Valenti-review}. 
Then the combinations of the Heisenberg interaction and the anisotropic exchange interactions
cause   
various types of magnetic order~\cite{Rau-PRL,Valenti-review,PD1,PD2,PD3,PD4,PD5};
if the Kitaev interaction is dominant, 
the spin-liquid states are stabilized~\cite{Kitaev}.
Controlling the exchange interactions via a time-periodic field
may provide a new opportunity to engineer their properties. 
Nevertheless, 
it is unclear how
a time-periodic field changes
the exchange interactions of Mott insulators with strong SOC.

In this work, 
we study the exchange interactions of
periodically driven Mott insulators with strong SOC.
We use a $t_{2g}$-orbital Hubbard model in the presence of strong SOC
and a circularly polarized light field on the honeycomb lattice
and derive its effective Hamiltonian in the strong-interaction limit.
Applying this theory to the case of $\alpha$-RuCl$_{3}$,
we show that
the magnitudes and signs of three exchange interactions
can be changed.
Then, studying another case of our model,
we show that
the Heisenberg interaction can be made much smaller than
the anisotropic exchange interactions.

\begin{figure}
  \includegraphics[width=50mm]{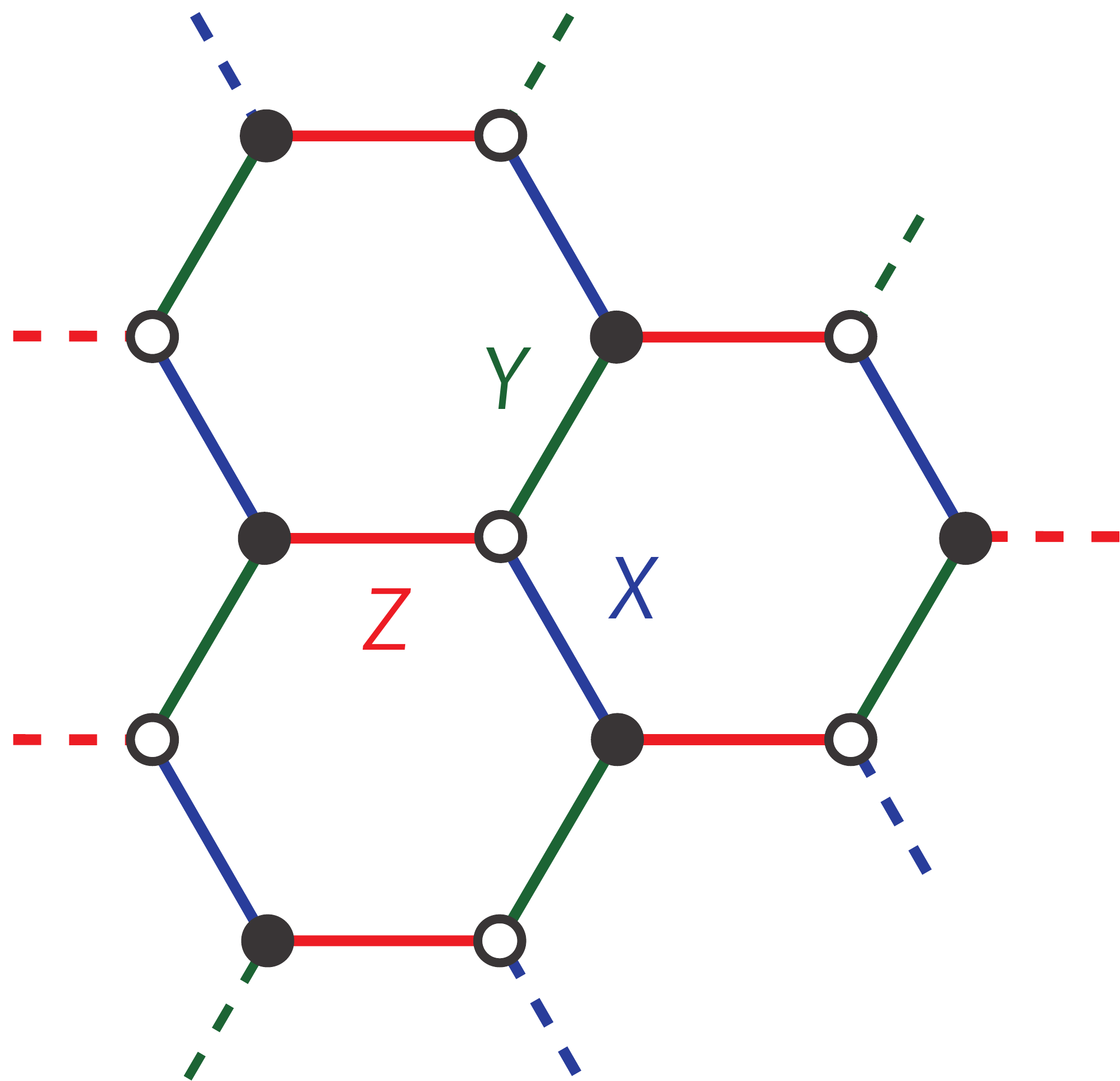}
  \caption{\label{fig1}
    Structure of the honeycomb lattice.
    $X$, $Y$, and $Z$ denote
    three different nearest-neighbor bonds.
    Black and white circles represent
    the $A$ and $B$ sublattices, respectively.
  }
\end{figure}

\textit{Setup. }We consider a periodically driven multiorbital system described by
\begin{align}
  H=H_{\textrm{KE}}+H_{\textrm{SOC}}+H_{\textrm{int}},
\end{align}
where $H_{\textrm{KE}}$, $H_{\textrm{SOC}}$, and $H_{\textrm{int}}$
represent the kinetic energy,
the atomic SOC~\cite{StrongLS-Mott-review},
and the local multiorbital Coulomb interactions~\cite{Kanamori},
respectively.
The kinetic energy is given by the hopping integrals of the $t_{2g}$-orbital
electrons on the honeycomb lattice (Fig. \ref{fig1})
in the presence of a circularly polarized light field
$\bdE(t)={}^{t}(E_{0}\cos\omega t\ -E_{0}\sin\omega t)$. 
The effects of $\bdE(t)$ are treated as Peierls phase factors: 
\begin{align}
  H_{\textrm{KE}}=
  \sum_{i,j}\sum_{a,b}\sum_{\sigma}
  t_{iajb}e^{-ie(\bdR_{i}-\bdR_{j})\cdot\bdA(t)}c_{ia\sigma}^{\dagger}c_{jb\sigma},
\end{align}
where $\bdA(t)={}^{t}(-\frac{E_{0}}{\omega}\sin\omega t\ -\frac{E_{0}}{\omega}\cos\omega t)$;
hereafter, we use $\hbar=1$. 
Then the atomic SOC of $H_{\textrm{SOC}}$ is given by
the $LS$ coupling for the $t_{2g}$-orbital electrons~\cite{StrongLS-Mott-review}.
The terms of $H_{\textrm{int}}$ consist of
the following interactions~\cite{Kanamori}:
\begin{align}
  &H_{\textrm{int}}=
  \sum_{i}\Bigl\{\sum_{a,b}
  c_{ia\uparrow}^{\dagger}c_{ia\downarrow}^{\dagger}
  [U\delta_{a,b}+J^{\prime}(1-\delta_{a,b})]
  c_{ib\downarrow}c_{ib\uparrow}\notag\\
  &+\sum_{\substack{a,b\\a>b}}\sum_{\sigma,\sigma^{\prime}}
  c_{ia\sigma}^{\dagger}c_{ib\sigma^{\prime}}^{\dagger}
  (U^{\prime}c_{ib\sigma^{\prime}}c_{ia\sigma}
  -J_{\textrm{H}}c_{ib\sigma}c_{ia\sigma^{\prime}})\Bigr\}.
\end{align}

As a specific example,
we consider a minimal model of $\alpha$-RuCl$_{3}$~\cite{Rau-PRL}:
$t_{iajb}$'s in $H_{\textrm{KE}}$ can be parametrized
by three hopping integrals between nearest-neighbor sites
on the honeycomb lattice (Fig. \ref{fig1}).
Namely, 
for the $Z$ bond,
the finite $t_{iajb}$'s are given by 
$t_{id_{yz}jd_{yz}}=t_{id_{zx}jd_{zx}}=t_{1}$,
$t_{id_{yz}jd_{zx}}=t_{id_{zx}jd_{yz}}=t_{2}$,
and $t_{id_{xy}jd_{xy}}=t_{3}$;
for the $X$ or $Y$ bond,
similar relations can be obtained from symmetry arguments~\cite{Rau-PRL}.
Then 
the low-energy properties can be described
by the $j_{\textrm{eff}}=1/2$ states~\cite{Rau-PRL,Jeff-RuCl3-1,Jeff-RuCl3-2}, 
$|+\rangle_{i}=\frac{1}{\sqrt{3}}
(c_{id_{yz}\downarrow}^{\dagger}+ic_{id_{zx}\downarrow}^{\dagger}+c_{id_{xy}\uparrow}^{\dagger})|0\rangle$
and
$|-\rangle_{i}=\frac{1}{\sqrt{3}}
(c_{id_{yz}\uparrow}^{\dagger}-ic_{id_{zx}\uparrow}^{\dagger}-c_{id_{xy}\downarrow}^{\dagger})|0\rangle$,
in which
the spin and the orbital are entangled by strong SOC.
Since such entanglement is the key property of
strong SOC~\cite{StrongLS-Mott-review},
this model will be sufficient for analyzing 
essential physics in the presence of strong SOC.

\textit{Floquet theory of Mott insulators. }We derive
an effective Hamiltonian for a periodically driven Mott insulator
using the Floquet theory~\cite{Floquet-MultiMott2}.
To derive it, 
we consider the strong-interaction limit in which  
$t_{iajb}$'s are much smaller than
the energies of doubly occupied states, 
$U+2J^{\prime}$, $U-J^{\prime}$, $U^{\prime}-J_{\textrm{H}}$,
and $U^{\prime}+J_{\textrm{H}}$~\cite{NA,Ishihara}.
In this limit,
we can approximately express the solution to the Schr\"{o}dinger equation,
$|\Psi\rangle_{t}$,
as $|\Psi\rangle_{t}\approx |\Psi_{0}\rangle_{t}+|\Psi_{1}\rangle_{t}$~\cite{Floquet-MultiMott2},
where $|\Psi_{0}\rangle_{t}$ and $|\Psi_{1}\rangle_{t}$
denote the states without and with, respectively, a doubly occupied site.
As a result,
the Schr\"{o}dinger equation reduces to a set of simultaneous equations:
\begin{align}
  &i\partial_{t}|\Psi_{0}\rangle_{t}
  =\mathcal{P}_{0}H_{\textrm{KE}}|\Psi_{1}\rangle_{t}
  +H_{\textrm{SOC}}|\Psi_{0}\rangle_{t},\label{eq:Psi0}\\
  &i\partial_{t}|\Psi_{1}\rangle_{t}
  =H_{\textrm{KE}}|\Psi_{0}\rangle_{t}
  +(\mathcal{P}_{1}H_{\textrm{KE}}\mathcal{P}_{1}
  +\tilde{H}_{\textrm{int}})|\Psi_{1}\rangle_{t},\label{eq:Psi1}
\end{align}
where $\tilde{H}_{\textrm{int}}=H_{\textrm{int}}+H_{\textrm{SOC}}$; 
and $\mathcal{P}_{0}$ and $\mathcal{P}_{1}$
denote the projections onto the subspaces without and with, respectively,
a doubly occupied site.
Hereafter, we concentrate on the high-frequency case in which
$\omega$ is much larger than $t_{iajb}$'s. 
In this case,
we could replace
the time-dependent operator $\mathcal{P}_{1}H_{\textrm{KE}}\mathcal{P}_{1}$
in Eq. (\ref{eq:Psi1})
by its time-averaged one
$\bar{H}_{\textrm{KE}}=\sum_{i,j}\sum_{a,b}\sum_{\sigma}t_{iajb}\mathcal{J}_{0}(u_{ij})
\mathcal{P}_{1}c_{ia\sigma}^{\dagger}c_{jb\sigma}\mathcal{P}_{1}$~\cite{Floquet-MultiMott2},
where $\mathcal{J}_{n}(u_{ij})$ is the $n$th Bessel function of the first kind
and $u_{ij}=\frac{eE_{0}}{\omega}\textrm{sgn}(i-j)$~\cite{remark1};
the distance between nearest-neighbor sites is set to unity. 
By using this replacement,
we can solve Eq. (\ref{eq:Psi1});
the result~\cite{Supp} is
\begin{align}
  |\Psi_{1}\rangle_{t}=
  \sum_{i,j,a,b,\sigma}\sum_{n=-\infty}^{\infty}
  \frac{t_{iajb}\tilde{\mathcal{J}}_{-n}(u_{ij})e^{-in\omega t}}
       {n\omega-\bar{H}_{\textrm{KE}}-\tilde{H}_{\textrm{int}}}
       c_{ia\sigma}^{\dagger}c_{jb\sigma}|\Psi_{0}\rangle_{t},\label{eq:Psi1-next}
\end{align}
where
$\tilde{\mathcal{J}}_{-n}(u_{ij})=\mathcal{J}_{-n}(u_{ij})e^{-in\theta_{ij}}$,
and $\theta_{ij}=\theta_{ji}=\frac{\pi}{3}$, $\pi$, and $\frac{5\pi}{3}$
for the $Y$, $Z$, and $X$ bonds, respectively.
Furthermore,
since
$H_{\textrm{int}}$ gives the largest contribution of
the terms of $\bar{H}_{\textrm{KE}}$ and $\tilde{H}_{\textrm{int}}(=H_{\textrm{int}}+H_{\textrm{SOC}})$, 
we replace $n\omega-\bar{H}_{\textrm{KE}}-\tilde{H}_{\textrm{int}}$ 
in Eq. (\ref{eq:Psi1-next}) by $n\omega-H_{\textrm{int}}$;
this replacement may be sufficient
if $\omega$ is nonresonant, i.e.,
the denominator of Eq. (\ref{eq:Psi1-next}) does not diverge.
By using Eq. (\ref{eq:Psi1-next}) with this replacement
and omitting the constant term (i.e., $H_{\textrm{SOC}}|\Psi_{0}\rangle_{t}$), 
we can rewrite Eq. (\ref{eq:Psi0}) as
\begin{align}
  i\partial_{t}|\Psi_{0}\rangle_{t}=H_{\textrm{eff}}(t)|\Psi_{0}\rangle_{t},\label{eq:Psi0-final}
\end{align}
where
\begin{align}
  H_{\textrm{eff}}(t)=&
  \sum_{i,j}\sum_{a,b,c,d}\sum_{\sigma,\sigma^{\prime}}
  \sum_{n,m=-\infty}^{\infty} 
  t_{jcid}t_{iajb}
  \mathcal{P}_{0}c_{jc\sigma^{\prime}}^{\dagger}c_{id\sigma^{\prime}}
  \notag\\
  \times &
  \frac{\tilde{\mathcal{J}}_{m}(u_{ji})\tilde{\mathcal{J}}_{-n}(u_{ij})e^{i(m-n)\omega t}}
       {n\omega-H_{\textrm{int}}} 
  c_{ia\sigma}^{\dagger}c_{jb\sigma}\mathcal{P}_{0}.\label{eq:Heff-t}
\end{align}

The leading term of $H_{\textrm{eff}}(t)$
is given by the time-independent Floquet Hamiltonian.
Since $H_{\textrm{eff}}(t)$ is time periodic,
it can be expressed as the Fourier series 
$H_{\textrm{eff}}(t)=\sum_{l}e^{il\omega t}H_{l}$.  
Furthermore,
by using a high-frequency expansion of the Floquet theory~\cite{review1,review2,review3},
$H_{\textrm{eff}}(t)$ can be written in the form
$H_{\textrm{eff}}(t)=H_{0}+O(\omega^{-1})$.
Therefore, 
the time-averaged $H_{\textrm{eff}}(t)$, $\bar{H}_{\textrm{eff}}$,
gives the leading term of Eq. (\ref{eq:Heff-t});
$\bar{H}_{\textrm{eff}}$ is given by
\begin{align}
  \bar{H}_{\textrm{eff}}
  =&\sum_{i,j}\sum_{a,b,c,d}\sum_{\sigma,\sigma^{\prime}}
  \sum_{n=-\infty}^{\infty}
  t_{jcid}t_{iajb}
  \mathcal{P}_{0}c_{jc\sigma^{\prime}}^{\dagger}c_{id\sigma^{\prime}}
  \frac{\mathcal{J}_{n}(u_{ij})^{2}}{n\omega-H_{\textrm{int}}}\notag\\
  &\times
  c_{ia\sigma}^{\dagger}c_{jb\sigma}\mathcal{P}_{0}.\label{eq:Floquet-H}
\end{align}

\begin{figure*}
  \includegraphics[width=178mm]{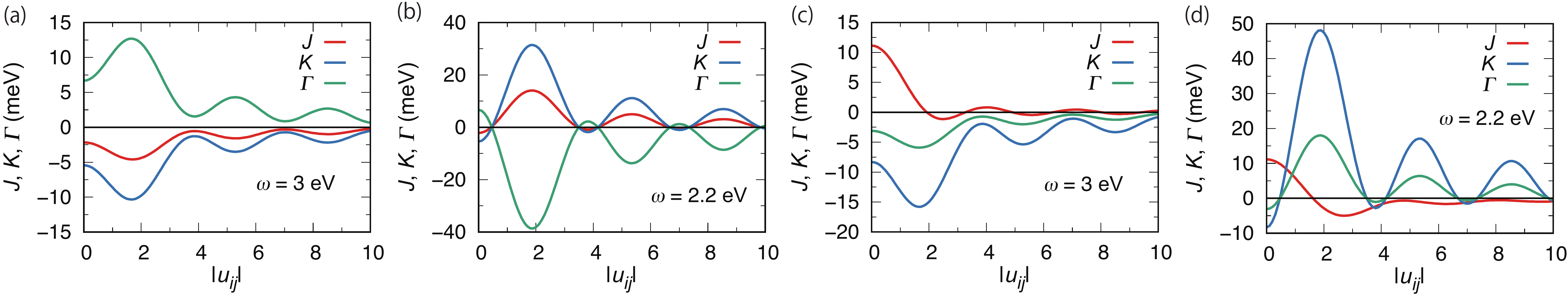}
  \caption{\label{fig2}
    The $|u_{ij}|(=|\frac{eE_{0}}{\omega}|)$ dependences of $J$, $K$, and $\Gamma$ 
    in (a) and (b) the case of $\alpha$-RuCl$_{3}$ 
    and (c) and (d) another case. 
  }
\end{figure*}

\textit{Application to periodically driven $\alpha$-RuCl$_{3}$. }Applying
the above theory to the minimal model of $\alpha$-RuCl$_{3}$,
we derive its Floquet Hamiltonian.
This derivation can be performed
in a way similar to the derivation in the absence of a driving field.
Here we describe the main points of the derivation 
(for the details, see the Supplemental Material~\cite{Supp}).
To derive the expression of $\bar{H}_{\textrm{eff}}$
for the minimal model of $\alpha$-RuCl$_{3}$,
we suppose that
in the subspace of $|\Psi_{0}\rangle_{t}$
a single hole occupies $j_{\textrm{eff}}=1/2$ state per site.
We also rewrite $H_{\textrm{int}}$
using the irreducible representations of doubly occupied states~\cite{NA}:
$H_{\textrm{int}}=\sum_{i}\sum_{\Gamma,g_{\Gamma}}U_{\Gamma}
|i;\Gamma,g_{\Gamma}\rangle\langle i;\Gamma,g_{\Gamma}|$,
where $U_{A_{1}}=U+2J^{\prime}$, $U_{E}=U-J^{\prime}$,
$U_{T_{1}}=U^{\prime}-J_{\textrm{H}}$, and $U_{T_{2}}=U^{\prime}+J_{\textrm{H}}$;
$|i;\Gamma,g_{\Gamma}\rangle$'s are expressed in the Supplemental Material~\cite{Supp}.
Then, by calculating the contributions of possible hopping processes to $\bar{H}_{\textrm{eff}}$,
we obtain~\cite{Supp}
\begin{align}
  \hspace{-10pt}
  \bar{H}_{\textrm{eff}}
  =\sum_{\langle i,j\rangle}
  [J\bdS_{i}\cdot\bdS_{j}
    +KS_{i}^{\gamma}S_{j}^{\gamma}
    +\Gamma(S_{i}^{\alpha}S_{j}^{\beta}+S_{i}^{\beta}S_{j}^{\alpha})],\label{eq:H-Floquet}
\end{align}
where
\begin{align}
  \hspace{-0pt}
  J
  =&\sum_{n=-\infty}^{\infty}\frac{4\mathcal{J}_{n}(u_{ij})^{2}}{27}
  \Bigl\{\frac{(2t_{1}+t_{3})^{2}}{U+2J^{\prime}-n\omega}
  +\frac{6t_{2}^{2}}{U^{\prime}+J_{\textrm{H}}-n\omega}\notag\\
  \hspace{-0pt}
  &\ \ \ \ \
  +\frac{2[(t_{1}-t_{3})^{2}-3t_{2}^{2}]}{U-J^{\prime}-n\omega}
  +\frac{6t_{1}(t_{1}+2t_{3})}{U^{\prime}-J_{\textrm{H}}-n\omega}\Bigr\},
  \label{eq:J}\\
  \hspace{-0pt}
  K
  =&\sum_{n=-\infty}^{\infty}\frac{4\mathcal{J}_{n}(u_{ij})^{2}}{9}
  \Bigl\{\frac{4t_{2}^{2}}{U-J^{\prime}-n\omega}
  -\frac{[(t_{1}-t_{3})^{2}+t_{2}^{2}]}{U^{\prime}+J_{\textrm{H}}-n\omega}\notag\\
  \hspace{-0pt}
  &\ \ \ \ \
  +\frac{[(t_{1}-t_{3})^{2}-3t_{2}^{2}]}{U^{\prime}-J_{\textrm{H}}-n\omega}\Bigr\},
  \label{eq:K}\\
  \hspace{-0pt}
  \Gamma
  =&\sum_{n=-\infty}^{\infty}\frac{16\mathcal{J}_{n}(u_{ij})^{2}t_{2}(t_{1}-t_{3})J_{\textrm{H}}}
  {9(U^{\prime}-J_{\textrm{H}}-n\omega)(U^{\prime}+J_{\textrm{H}}-n\omega)},
  \label{eq:Gam}
\end{align}
and $(\alpha,\beta,\gamma)=(x,y,z)$, $(y,z,x)$, and $(z,x,y)$
for the $Z$, $X$, and $Y$ bonds, respectively;
$J$, $K$, and $\Gamma$
are the Heisenberg interaction,
the Kitaev interaction,
and the off-diagonal symmetric exchange interaction,
respectively.
[These expressions hold also for 
$\bdE(t)={}^{t}(E_{0}\cos\omega t\ E_{0}\sin\omega t)$.]

We now show how
$J$, $K$, and $\Gamma$ vary with $\omega$ and $u_{ij}$.
To do it,
we numerically evaluate Eqs. (\ref{eq:J}){--}(\ref{eq:Gam}).
We set
$t_{1}=47$ meV, $t_{2}=160$ meV, $t_{3}=-129$ meV~\cite{remark2}, 
$J^{\prime}=J_{\textrm{H}}$, 
$U^{\prime}=U-2J_{\textrm{H}}$,
$U=3$ eV, and $J_{\textrm{H}}=0.5$ eV;
we replace $\sum_{n=-\infty}^{\infty}$'s by $\sum_{n=-n_{\textrm{max}}}^{n_{\textrm{max}}}$'s 
and set $n_{\textrm{max}}=500$. 
Figures \ref{fig2}(a) and \ref{fig2}(b) show
the $|u_{ij}|$ dependences of $J$, $K$, and $\Gamma$ 
at $\omega=3$ and $2.2$ eV.
We see that
by changing $|u_{ij}|$, 
the magnitudes of $J$, $K$, and $\Gamma$ can be changed
at $\omega=3$ and $2.2$ eV;
and that at $\omega=2.2$ eV
it is possible to change 
not only their magnitudes but also their signs.

For a deeper understanding of the above results,
we perform some analyses of Eqs. (\ref{eq:J}){--}(\ref{eq:Gam}).
Since $J^{\prime}=J_{\textrm{H}}$ and $U^{\prime}=U-2J_{\textrm{H}}$,
$J$, $K$, and $\Gamma$ can be rewritten as follows:
$J=J_{1}+J_{2}+J_{3}$, where
$J_{1}=\sum_{n}\frac{4\mathcal{J}_{n}(u_{ij})^{2}(2t_{1}+t_{3})^{2}}{27(U+2J_{\textrm{H}}-n\omega)}$,
$J_{2}=\sum_{n}\frac{8\mathcal{J}_{n}(u_{ij})^{2}(t_{1}-t_{3})^{2}}{27(U-J_{\textrm{H}}-n\omega)}$,
and $J_{3}=\sum_{n}\frac{8\mathcal{J}_{n}(u_{ij})^{2}t_{1}(t_{1}+2t_{3})}{9(U-3J_{\textrm{H}}-n\omega)}$.
$K=K_{1}+K_{2}$,
where
$K_{1}=\sum_{n}
\frac{4\mathcal{J}_{n}(u_{ij})^{2}[3t_{2}^{2}-(t_{1}-t_{3})^{2}]}{9(U-J_{\textrm{H}}-n\omega)}$and 
$K_{2}=\sum_{n}
\frac{4\mathcal{J}_{n}(u_{ij})^{2}[(t_{1}-t_{3})^{2}-3t_{2}^{2}]}{9(U-3J_{\textrm{H}}-n\omega)}$.
$\Gamma=\Gamma_{1}+\Gamma_{2}$,
where
$\Gamma_{1}=\sum_{n}\frac{8\mathcal{J}_{n}(u_{ij})^{2}t_{2}(t_{1}-t_{3})}{9(U-3J_{\textrm{H}}-n\omega)}$
and $\Gamma_{2}=\sum_{n}\frac{8\mathcal{J}_{n}(u_{ij})^{2}t_{2}(t_{3}-t_{1})}{9(U-J_{\textrm{H}}-n\omega)}$. 
For the hopping parameters of $\alpha$-RuCl$_{3}$,
$J_{1}$ is much smaller than $J_{2}$ and $J_{3}$;
as a result, $J\approx J_{2}+J_{3}$.
This is the origin of
the in-phase $|u_{ij}|$ dependences of $J$, $K$, and $\Gamma$ 
[Figs. \ref{fig2}(a) and \ref{fig2}(b)].
Then
we can understand 
the sign changes of $J$, $K$, and $\Gamma$
at $|u_{ij}|\sim 0.4$, $3.5$ [Fig. \ref{fig2}(b)]
by estimating the dominant contributions. 
We make the estimate of $J$ 
because the sign changes of $K$ and $\Gamma$
can be understood similarly.
For $\omega=2.2$ eV,
the dominant contributions are given by
\begin{align}
  J\approx
  (J_{2}^{0}+J_{3}^{0})\mathcal{J}_{0}(u_{ij})^{2}
  +(J_{2}^{0}c_{2}-J_{3}^{0}c_{3})\mathcal{J}_{1}(u_{ij})^{2},\label{eq:J-approx-w22}
\end{align}
where
$J_{2}^{0}=\frac{8(t_{1}-t_{3})^{2}}{27(U-J_{\textrm{H}})}$,
$J_{3}^{0}=\frac{8t_{1}(t_{1}+2t_{3})}{9(U-3J_{\textrm{H}})}$,
$c_{2}=\frac{U-J_{\textrm{H}}}{\delta \omega_{2}}$,
$c_{3}=\frac{U-3J_{\textrm{H}}}{\delta \omega_{3}}$,
and 
$\omega=U-3J_{\textrm{H}}+\delta \omega_{3}=U-J_{\textrm{H}}-\delta \omega_{2}$
(i.e., $\delta\omega_{2}=0.3$ eV and $\delta\omega_{3}=0.7$ eV). 
At $|u_{ij}|=0$,
$J$ is ferromagnetic, i.e., negative,
because
$J_{2}^{0}$ and $J_{3}^{0}$ satisfy 
$J_{2}^{0}>0$, $J_{3}^{0}<0$, and $J_{2}^{0}+J_{3}^{0} < 0$.  
As $|u_{ij}|$ increases,
the term including $\mathcal{J}_{1}(u_{ij})^{2}$ in Eq. (\ref{eq:J-approx-w22}),
the positive-sign contribution, becomes considerable
and causes a sign change of $J$. 
With further increases in $|u_{ij}|$,
$\mathcal{J}_{1}(u_{ij})^{2}$ approaches zero,
and the sign of $J$ changes again.

\textit{Application to another case. }We consider
another case and study the effects of the driving field
on the exchange interactions. 
In this case,
we set $t_{3}=129$ meV
and use the same values of the other parameters
as those used in the case of $\alpha$-RuCl$_{3}$;
in a set of these values,
$J_{1}$ is comparable to $J_{2}$ and $J_{3}$.
Although it may be difficult to change the value of $t_{3}$
in $\alpha$-RuCl$_{3}$,
we study this case to clarify 
how the driving field changes $J$  
in the presence of non-negligible $J_{1}$.
Figures \ref{fig2}(c) and \ref{fig2}(d)
show the $|u_{ij}|$ dependences of $J$, $K$, and $\Gamma$
in this additional case.
We see that
the $|u_{ij}|$ dependence of $J$ differs from that of $K$ or $\Gamma$.
In particular,
$J$ can be very small in magnitude,
while $K$ and $\Gamma$ are finite
[see, for example, their values at $|u_{ij}|=1.6$
in Fig. \ref{fig2} (d)].

\textit{Discussion. }We comment on the validity of our theory.
First, the hopping integrals of our model are simplified
compared with those obtained in the first-principles calculations~\cite{Valenti-PRB}.
However, since the leading terms are $t_{2}$ and $t_{3}$~\cite{Valenti-PRB},
our model may be appropriate for a minimal model of $\alpha$-RuCl$_{3}$.
Then
the obtained $|u_{ij}|$ dependences of $J$, $K$, and $\Gamma$
might be affected by
the doublon-holon hoppings
described by $\bar{H}_{\textrm{KE}}$.
Nevertheless,
we believe our results remain qualitatively unchanged.
This is 
because
the previous studies~\cite{Floquet-MultiMott1,Floquet-MultiMott2} show that
in the frequency range where 
the correction due to $\mathcal{J}_{1}(u_{ij})^{2}$ is important
and the corrections due to $\mathcal{J}_{n}(u_{ij})^{2}$'s for $n\geq 2$ are less important
(the range of $U-2J_{\textrm{H}}<\omega< U$ in Ref. \onlinecite{Floquet-MultiMott1}),
the effects of the driving field on the exchange interactions
remain qualitatively unchanged 
even if the doublon-holon hoppings are taken into account.

We also remark on heating effects.
The periodically driven system eventually approaches
an infinite-temperature state~\cite{Heat-InfT1,Heat-InfT2}.
However, at intermediate times $t\lesssim \tau$~\cite{remark3},
it can be approximately described by
the Floquet Hamiltonian
as long as $\omega$ is nonresonant~\cite{Floquet-MultiMott2}
and much larger than
the exchange interactions~\cite{Kuwahara,Heat-Floq1,Heat-Floq2,Heat-Floq3,Heat-Floq4}.
Since these conditions hold in our study, 
the properties similar to our results
could be realized experimentally. 

We now discuss the implications of our results.
First,
our results in the case of $\alpha$-RuCl$_{3}$
indicate that by tuning $\omega$ and
changing $E_{0}$,
one can change the magnitudes and signs of
the three exchange interactions of periodically driven $\alpha$-RuCl$_{3}$.
In particular,
by using this method, 
the Kitaev interaction can be made
ferromagnetic (negative) or antiferromagnetic (positive).
Since its sign drastically affects the magnetic properties of
materials with strong SOC~\cite{Valenti-review,Vojta-review},
our results will provide an opportunity for connecting
the ferromagnetic and the antiferromagnetic Kitaev physics.
Such control of the exchange interactions could be achieved
by pump-probe measurements.
Then
our results in another case
suggest that
if the contribution from the doubly occupied state with $A_{1}$ symmetry
is non-negligible,
it is possible to 
make the Kitaev interaction 
much larger in magnitude than the Heisenberg interaction.
Therefore
the periodically driven Mott insulator with strong SOC and
hopping integrals that lead to such a contribution
may be suitable for realizing the Kitaev model~\cite{Kitaev} and the spin liquid.

\textit{Conclusions. ]}We have studied the exchange interactions
of the Mott insulators with a circularly polarized light field
and strong SOC in two cases.
In the case of $\alpha$-RuCl$_{3}$,
we have shown that
$J$, $K$, and $\Gamma$
have the similar $|u_{ij}|$ dependences,
and that
their magnitudes and signs can be changed by tuning $\omega$ and varying $E_{0}$.
These properties can be utilized 
for changing the exchange interactions of $\alpha$-RuCl$_{3}$
and controlling its magnetic properties. 
In another case,
we have shown that
the $|u_{ij}|$ dependence of $J$ differs from those of $K$ and $\Gamma$,
and that
$J$ can be made much smaller than $K$ and $\Gamma$
by tuning $|u_{ij}|$.
The latter property suggests a new possibility
of realizing the Kitaev spin liquid.
Our results will provide the first step towards
controlling 
the exchange interactions and magnetic properties
of periodically driven Mott insulators with strong SOC. 

\begin{acknowledgments}
  This work was supported by
  JST CREST Grant No. JPMJCR1901, 
  JSPS KAKENHI Grants No. JP19K14664 and No. JP16K05459,
  and MEXT Q-LEAP Grant No. JP-MXS0118067426.
\end{acknowledgments}


\begin{references}
\bibitem{remark1}
  $\textrm{sgn}(i-j)=1$ and $-1$ 
  when $i$ is on the $A$ and $B$ sublattices (Fig. \ref{fig1}), respectively.
\bibitem{remark2}
  Our $t_{1}$, $t_{2}$, and $t_{3}$ correspond to the averages of
  $t_{1}$, $t_{1a}^{\prime}$, and $t_{1b}^{\prime}$ of Ref. \onlinecite{Valenti-PRB},
  of $t_{2}$ and $t_{2}^{\prime}$,
  and of $t_{3}$ and $t_{3}^{\prime}$, respectively.
\bibitem{remark3}
  $\tau$ may be sufficiently large
  because it is roughly given by 
  $\tau\approx T\exp(\omega/J_{\textrm{ex}})$~\cite{Kuwahara}
  and our $\omega$ and exchange interactions satisfy
  $\omega=O(1\textrm{eV})$ and
  $J_{\textrm{ex}}=\textrm{max}(J,K,\Gamma)=O(10\textrm{meV})$.
\end{references}
\end{document}